\ifx\pdfminorversion\undefined\else\pdfminorversion=7\fi
\ifx\texorpdfstring\undefined\newcommand\texorpdfstring[2]{{#1}}\fi
\documentclass[aps,prd,12pt,onecolumn,groupedaddress,tightenlines,nofootinbib,preprintnumbers,floatfix]{revtex4}%

\usepackage{enumitem}

\usepackage{subfigure} 
\ifnum\paperheight=0\relax\paperheight=11in\fi
\usepackage{amssymb,amsmath,graphicx,xcolor,cancel,slashed}

\usepackage{amsfonts} 
\usepackage{subfigure} 
\usepackage{array} 
\usepackage{dcolumn} 
\usepackage{bm} 
\usepackage{latexsym} 
\usepackage{longtable} 
\usepackage{bbold}
\usepackage{color}
\usepackage{multirow}
\usepackage{rotating}
\usepackage{comment}

\newcommand{\CPV}{$\cancel{\text{CP}}$}

\makeatletter
\let\@bibitemShut\relax
\makeatother

\usepackage[hyperfootnotes=false]{hyperref}
\begin{document}

\title{SYNERGY BETWEEN NP AND HEP RESEARCH GOALS AND EFFORTS IN FUNDAMENTAL SYMMETRIES AND INTERACTIONS}
\date{\today}
\author{Tanmoy Bhattacharya}
\affiliation{Los Alamos National Laboratory, T-2, Los Alamos, NM 87545, USA}
\author{Rajan Gupta}
\affiliation{Los Alamos National Laboratory, T-2, Los Alamos, NM 87545, USA}
\author{Kate Scholberg}
\affiliation{Department of Physics, Duke  University, Durham, NC, 27708, USA}
\begin{abstract}
The aim of this white paper is to highlight several areas for which the Department of Energy's Office of Nuclear Physics has primary stewardship or significant investment and expertise, and for which there is also significant interest and expertise within the HEP community.  These areas of overlap offer exciting opportunities for collaboration.
\end{abstract}
\maketitle


\section*{}
\label{sec:intro}
\vspace{-0.1in}
The 2021 Snowmass process brought to the fore a remarkable collaboration
between nuclear and high energy physicists
to elucidate the potential for significant progress through joint
experimental and theoretical efforts in four areas of great interest to
the ``Fundamental Symmetries'' subprogram of the DOE Office of Science, Nuclear
Physics. This collaboration is evident from the joint authorship of four contributions~\cite{USQCD:2022mmc,Ruso:2022qes,Alarcon:2022ero,Cirigliano:2022oqy}, including the associated topical group reports \cite{Davoudi:2022bnl,deGouvea:2022gut,Blum:2022cie}.
These four areas are: (i) neutrinoless double beta decay
($0\nu \beta \beta$), (ii) the neutron electric dipole moment (nEDM), 
(iii) tests of CKM unitarity through precision calculations for the extraction of the $V_{ud}$
matrix element, and (iv) lepton-nucleus scattering.  
 In addition, there are ongoing searches for novel scalar and tensor interactions at the TeV scale, and $N \overline N$ oscillations for baryon number violation. Conclusive results in any of these areas could merit the Nobel prize, and
will open new directions in beyond-the-standard-model (BSM) physics. In this short document we summarize the physics goals, the open
challenges and why collaborative efforts by multiple communities would 
greatly accelerate progress.\footnote{We note that the areas highlighted here do not represent all possible opportunities for joint NP/HEP collaboration.  For example, instrumentation development challenges are shared between the communities as well.}

{{\textbf {Neutrinoless double beta decay}}~\cite{Jones:2021cga}}: 
A signal in experiments searching for $0\nu \beta \beta$ will be a clear evidence of lepton-number-violation (LNV) and will demonstrate the Majorana nature of neutrinos. 
An observation in the next-generation experiments will either identify the neutrino mass ordering or, if oscillation experiments and 
advances in cosmology will show that neutrinos are organized in the ``normal ordering'', 
may provide decisive evidence of BSM physics, shedding light on the mechanism of neutrino mass generation.  
There are several experiments worldwide~\cite{Zerobb},
with the US program stewarded by DOE NP pursuing a multi-experiment international strategy.

$0\nu\beta\beta$ experiments are sensitive to a variety of LNV mechanisms, from the ``standard mechanism'' of light-Majorana-neutrino exchange, to contributions mediated by new particles at the TeV scale, or by weakly coupled light particles such as sterile neutrinos. Identifying the microscopic mechanism behind a signal demands a rich 
theoretical program over a wide range of energy scales \cite{Cirigliano:2022oqy}. At high energy, particle physics models with LNV need to be further developed, and the complementarity between $0\nu\beta\beta$ experiments, cosmology, and searches at present and future high-energy colliders needs to be further explored.

The $0\nu\beta\beta$ rates induced by light-Majorana exchange or less minimal LNV models can be computed by using a tower of effective field theories (EFTs), systematically linking the electroweak to the nuclear scale.
Because of the lack of experimental data, the couplings in the nuclear EFTs
need to be determined directly from QCD. 
Lattice QCD is currently the only way to systematically and reliably compute the necessary matrix elements. Significant progress has already been achieved in the calculation of LNV pion couplings
\cite{Nicholson:2018mwc,Detmold:2022jwu,Detmold:2020jqv,Tuo:2019bue,Tuo:2022hft}.
The determination of $0\nu\beta\beta$
transition operators requires, in addition,  nucleon-nucleon LNV couplings \cite{Cirigliano:2018yza}, even for  light-Majorana-neutrino exchange \cite{Cirigliano:2018hja}. 
Progress on this front will require further theoretical developments to relate lattice QCD results to physical two-nucleon matrix elements~\cite{Davoudi:2020gxs,Davoudi:2021noh},
coupled with computational advances 
to obtain precise two-nucleon spectra and matrix elements.

The results from Lattice QCD will then serve as input for many-body calculations of nuclear matrix elements (NME) in experimentally relevant isotopes.
Here \textit{ab initio} methods are
starting to appear alongside more traditional phenomenological approaches.
If accompanied by more Lattice QCD and EFT work
towards the construction of nuclear interactions and transition operators at the same order and in the same regularization scheme,
these methods will provide NMEs, and thus $0\nu\beta\beta$ rates, with a controlled estimate of the  theoretical uncertainties.

{{\textbf {Neutron Electric Dipole Moment
(nEDM)}}~\cite{Alarcon:2022ero}}: One of the profound mysteries of
nature is the lack of matter-antimatter symmetry in the universe,
i.e., the almost total absence of antibaryons. The symmetry between
baryons and antibaryons is expected to have been broken during the
evolution of the universe post inflation~\cite{Coppi:2004za}, and
requires CP violation (\CPV)~\cite{Sakharov:1967dj}. If it is in the quark sector, then it
has to be larger than that present in the CKM quark mixing
matrix~\cite{Sakharov:1967dj}. In that case, weak-scale Baryogenesis is the favored mechanism for
creating the asymmetry~\cite{Barrow:1981bv}. If it is in the
neutrino mixing matrix, then it would be through
Leptogenesis~\cite{FUKUGITA198645}. Any \CPV\ interaction in the quark sector necessarily contributes to the nEDM, and most popular BSM models have
additional \CPV\ that would give a $d_n > 10^{-28}$ e-cm~\cite{Pospelov:2005pr}.

The DOE NP Flagship SNS EDM experiment being built in the US at Oak
Ridge is designed to reach $d_n \sim 3 \times 10^{-28}$ e-cm~\cite{SNSEDM}, and there is a
less ambitious effort at LANL~\cite{LANLEDM} using already proven technology.  A successful
measurement will 
give credence to electroweak
baryogenesis~\cite{Morrissey:2012db} as the mechanism for the baryon asymmetry. The value (or the lowering of the 
bound in case of a null result) for $d_n$ will provide stringent constraints on possible BSM
theories, provided results for the matrix elements of low energy novel
\CPV\ operators of dimension six or less can be calculated between the neutron
ground state with $O(20\%)$ accuracy.  
Lattice QCD~\cite{Alarcon:2022ero}, with
effective field theory methods~\cite{Burges:1983zg} providing the connection
between \CPV\ couplings in BSM theories and the low-energy
effective \CPV\ operators~\cite{Pospelov:2005pr,Engel:2013lsa,Chupp:2017rkp}, is attempting to reach this precision over the next decade---there are currently multiple collaborations between nuclear
and HEP physicists doing the lattice and the EFT calculations to achieve this. This
combined effort is designed to elucidate fundamental symmetries and interactions at far beyond the TeV scale, often complementary to the searches at the LHC.

{{\textbf {Lepton-Nucleus scattering}}~\cite{Ruso:2022qes}}: The
flagship of the HEP program in the US is the DUNE experiment at
Fermilab~\cite{DUNE}. It is designed to quantify \CPV\ in the
neutrino sector. Since there is \CPV\ in the quark sector, it is important to quantify it in the neutrino sector. 
Reaching the design precision requires accurate measurements of the
$\nu$-nucleus cross-section. Essential, but the least constrained,
ingredients for this are the nucleon axial vector form factors and transition matrix elements over the
range of a few hundred MeV to a couple of GeV incident neutrino energy, and
corrections to these from nuclear effects~\cite{Ruso:2022qes}. This energy range covers the difficult-to-model
quasi-elastic and resonant regions, making the cross-section
calculations and Monte Carlo event generators challenging. The most promising approach to reach the required precision is to use lattice QCD to calculate the axial form factors of the nucleons and input them into nuclear many-body calculations of the cross-section.  \looseness-1

At lower energies (few to few-hundred MeV),  neutrino-nucleus interactions are relevant for astrophysical neutrinos (e.g., solar, atmospheric and supernova neutrinos), and their understanding is important both for the interpretation of detected signals and for processes occurring in the sources.  Thus, astrophysical signals provide information on both the sources and the properties of neutrinos themselves.  Neutrino cross-section measurements in this regime are also relevant for the understanding of weak couplings and nuclear transitions, as well as for searches for BSM physics~\cite{Huber:2022lpm,Balantekin:2022jrq}.  
Experimental data in this energy regime are sparse and theoretical understanding is also modest.  Joint HEP-NP efforts for both theory and experiment are underway, for example in the context of experiments at stopped-pion sources~\cite{Akimov:2022oyb,Asaadi:2022ojm,VandeWater:2022qot}.

The planned electron-ion collider (EIC) is designed to provide a detailed 3D tomographic map of the structure of nucleons in terms of quarks and gluons~\cite{AbdulKhalek:2021gbh}. Experiments at the EIC will significantly improve the measurements of electric and magnetic form factors that also enter the analysis of $\nu$-nucleus interactions. Similarly, improvements in the extraction of parton distribution functions are of interest to both the NP and HEP communities~\cite{Lin:2017snn}.

In all three areas, the ongoing
collaborative efforts between HEP and NP physicists again demonstrate that the relevant communities are already working together.

{{\textbf {Test of CKM unitarity}}~\cite{Workman:2022ynf,Seng:2018yzq,Seng:2018qru,Czarnecki:2019mwq}}:
Understanding of nuclear $\beta$ decays was instrumental in the discovery of the Standard Model. 
Even in the era of the LHC, $\beta$ decay experiments can probe BSM physics at scales of $\gtrsim 10$ TeV, highly competitive with direct searches. Tests of unitarity of the first row of the Cabibbo-Kobayashi-Maskawa mixing matrix are particularly sensitive to these effects. Recently, 
a  revaluation of
the ``inner radiative correction'' \cite{Seng:2018yzq, Czarnecki:2019mwq,Shiells:2020fqp,Hardy:2020qwl}
has led to a reduction of the uncertainty in the extraction of $V_{ud}$ from superallowed $0^+ \rightarrow 0^+$ decays, while 
progress in lattice QCD resulted in 
permille accuracy on the form factor $f_+(0)$
and on the ratio $f_{K^+}/f_{\pi^+}$,
needed to  extract  $V_{us}$
and $V_{us}/V_{ud}$ from kaon decays \cite{FlavourLatticeAveragingGroupFLAG:2021npn}.
These advances revealed a $\sim 3\sigma$ tension with the SM \cite{Workman:2022ynf,Seng:2018yzq,Seng:2018qru,Czarnecki:2019mwq}. 
Understanding the tension is limited by theoretical errors, with an uncertainty currently dominated by nuclear corrections in $0^+ \rightarrow 0^+$ decays~\cite{Hardy:2020qwl}.
In the near future, measurements of the neutron lifetime $\tau_n$ 
with uncertainty  $\Delta\tau_n \sim 0.1$~s, 
and of ratio $\lambda = g_A/g_V$ of the neutron axial and vector coupling 
with uncertainty $\Delta\lambda/|\lambda|\sim 0.03\%$, will allow for the extraction of $V_{ud}$ from neutron decay with accuracy comparable to superallowed $\beta$ decay. Such an extraction will have the advantage of not being affected by nuclear corrections.
Lattice QCD can play an important role in validating and reducing the error on 
the radiative corrections to meson and nucleon decays. The first calculations for pion and kaon decays have already appeared  \cite{Feng:2020zdc,Seng:2020wjq,Seng:2020jtz,Ma:2021azh,Yoo:2022lmt}, and work on nucleon decay is ongoing.
In addition to CKM unitarity, decay spectra and correlations also provide tests of new charged-current interactions at scales of about 10 TeV. Lattice QCD has provided precise calculations of the scalar and tensor charges
\cite{Gupta:2018qil,Alexandrou:2019brg,Liu:2021irg,Park:2021ypf,FlavourLatticeAveragingGroupFLAG:2021npn}, which are needed to convert bounds on the Fierz interference terms onto bounds on quark-level operators (see below).
Comparing experimental extractions and lattice QCD calculations of 
the nucleon axial charge $g_A$ can provide strong bounds on right-handed charged currents.
With lattice QCD approaching the percent level precision~\cite{Chang:2018uxx,Gupta:2018qil,Aoki:2019cca,FlavourLatticeAveragingGroupFLAG:2021npn}, these comparisons are now limited by electromagnetic corrections
\cite{Cirigliano:2022hob}.\looseness-1

\vspace{0.2in}

{{\textbf {Novel Scalar and Tensor Interactions at the TeV scale}}~\cite{Gonzalez-Alonso:2018omy}}:  The two communities are also working to search for 
novel scalar and tensor interactions at the TeV scale. 
The low-energy approach requires precision measurements of the neutron or nuclear decay distributions, the calculation of neutron matrix elements using lattice QCD~\cite{Bhattacharya:2011qm,Gupta:2018qil},
and, in the case of nuclear decays, the $\textit{ab initio}$ calculation of nuclear matrix elements.
At the moment, the best bounds on the neutron Fierz interference term, a probe of both scalar and tensor currents, come from the UCNA and Perkeo III experiments~\cite{Saul:2019qnp,UCNA:2019dlk},
while the Nab experiment will provide bounds of a few per-mil~\cite{Fry19}. The Fierz interference term induced by scalar interactions is sensitively probed in $0^+ \rightarrow 0^+$ superallowed $\beta$ decays \cite{Hardy:2020qwl}, while new experiments such as He6-CRES \cite{Cirigliano:2019wao,Garcia}
can investigate TeV-scale tensor currents.
At high energy,
scalar and tensor interactions affect the high transverse mass tail of the charged-current Drell-Yan process at the LHC \cite{Cirigliano:2012ab}.
The latest high-transverse-mass Drell-Yan dataset
from the ATLAS and CMS collaborations
\cite{ATLAS:2019lsy,CMS:2022yjm}, 
which use the full luminosity of the LHC Run II,
provide constraints on scalar and tensor interactions that are very competitive with present and future $\beta$ decay experiments \cite{Allwicher:2022mcg}.

{\textbf {$\Delta B =2$ baryon number violation in $N {\overline N}$ oscillations}}:
The current limit on the free neutron oscillation time $\tau_{N \overline N} \gtrsim 10^8$ sec can be converted into new physics scales of $10^2-10^3$ TeV, and upcoming experiments at the European Spallation Source will probe parameter space relevant to low-scale baryogenesis scenarios in which the baryon asymmetry is induced by the B violating decays of new particles that mediate ${N\overline N}$ oscillations~\cite{Babu:2006xc,Grojean:2018fus}. 

{\textbf {Synergy in theoretical methods used}}: There is close synergy and often collaborations between NP and HEP physicists exploiting two theoretical tools  needed to achieve physics goals: lattice QCD and effective field theory methods. 

{{\textbf {Lattice QCD}}~\cite{USQCD:2022mmc,Aoki:2019cca,FlavourLatticeAveragingGroupFLAG:2021npn}}:
Large-scale simulations of lattice QCD is the most promising tool for many of the theoretical calculations of matrix elements
needed in all  physics drivers. Explicit examples are connecting the $0\nu \beta \beta$, nEDM, neutron decay distributions, and $N \bar N$ oscillation experiments to BSM physics, and obtaining crucial input in the the
extraction of $V_{ud}$, $V_{us}$ and axial vector form factors for
lepton-nucleus scattering.
The US lattice QCD communities in both nuclear and high energy physics
collaborate and work jointly, for example, to procure resources that
are then allocated by the umbrella USQCD
collaboration~\cite{USQCD}. Many of the teams that receive these awards have members from both
communities working collaboratively on the above six areas and have a history of producing state-of-the-art results. These efforts would benefit from an increase in computing resources.

{\textbf {Effective Field Theory Methods}}~\cite{Buchmuller:1985jz} EFT is a systematic method to express interactions and their couplings arising in BSM theories in terms of low-energy effective operators composed of quark and gluon fields and organized by symmetries and dimension (roughly translating into importance). The renomalization group and QCD perturbation theory are used to run the associated couplings from the high scale to the hadronic scale of a few GeV, and in the process integrating out the heavy degrees of freedom systematically. Lattice QCD can then be used to calculate, incorporating full non-perturbative QCD dynamics, the matrix elements of these effective operators between hadron states. These matrix elements then provide the connection between low energy experiments and possible fundamental theories, for example, between bound/value of neutron EDM and allowed values for \CPV\ couplings in BSM theories, i.e., constraining the space of possible theories.

\acknowledgments

T. Bhattacharya and R. Gupta,  were partly supported by the U.S. DOE, Office of Science, HEP under Contract
No. DE-AC52-06NA25396 and the LANL LDRD program.   K. Scholberg is funded by the Department of Energy Office of Science, HEP and the National Science Foundation.

\pagebreak
\let\url\relax \let\Url\relax
\def\stringify#1{\global\def#1{\string#1\discretionary{}{}{}}}
{\catcode`\/=\active\stringify/\catcode`\_=\active\global\def_{\protect\textunderscore\discretionary{}{}{}}%
\catcode`\.=\active\stringify.%
\global\def\url#1{{\def\discretionary##1##2##3{}\let\protect\noexpand\edef\tmp{\noexpand\href{#1}}\expandafter}\tmp{#1}}%
\global\def\doi#1{{\def\discretionary##1##2##3{}\let\protect\noexpand\edef\tmp{\noexpand\href{https://doi.org/#1}}\expandafter}\tmp{doi:#1}}%
}
\def\Url{{\catcode`\/=\active\catcode`\_=\active\catcode`\.=\active\expandafter}\url}
\def\Doi{{\catcode`\/=\active\catcode`\_=\active\catcode`\.=\active\expandafter}\doi}
%
\bibliographystyle{h-physrev-doi} 
\bibliography{ref,nEDM,FF,solvers,ref1,ref2,lqcd} 

\end{document}